\documentclass[%
 reprint,
 amsmath,amssymb,
pra,
]{revtex4-1}

\usepackage{graphicx}
\usepackage{dcolumn}
\usepackage{bm}
\usepackage{ulem}

\begin{document}

\preprint{APS/123-QED}
\title{Scattering of partially coherent radiation by non-Hermitian localized structures having Parity-Time symmetry}

\author{P. A. Brand\~ao}
\email{paulo.brandao@fis.ufal.br}
\author{S. B. Cavalcanti}%
\email{solange@fis.ufal.br}
\affiliation{%
 Universidade Federal de Alagoas, Instituto de F\'isica, 57072-900, Brazil
}%




\date{\today}

\begin{abstract}
A theoretical model based on two point scatterers is suggested to investigate for the first time scattering of partially coherent radiation by a non-Hermitian localized structure, invariant under the simultaneous symmetry operations of parity inversion and time reversal. Within the first-order Born approximation, and the formalism of classical coherence theory, the spectral density of the scattered field in the far-zone is obtained analytically. We find that the unidirectional character of the scattered radiation, which is one of the principal effects present in non-Hermitian structures, may be suppressed due to the coherence properties of the radiation without changing the geometry of the structure. Asymmetric spectral changes of the scattered radiation due to the non-Hermitian character of the material is also observed. 
\end{abstract}

\maketitle


\section{\label{sec:intro} Introduction}

Since the seminal work of Bender and Boettcher on the existence of all-real eigenvalue spectra exhibited by non-Hermitian Hamiltonians, work on the so-called Parity-Time (PT)-symmetric structures have been pursued worldwide, with unexpected and interesting results, opening a new and exciting way of interpreting the consequences of crossing to the forbidden complex domain \cite{Bender1998a,Bender1999,Bender2002,Bendermust,Bendermakingsense,Benderbook}.  A PT-symmetric quantum system is characterized by a potential function invariant under the simultaneous symmetry operations of parity inversion (P) and time reversal (T). Considering the condition for a generic Hamiltonian $H$ to be PT-symmetric, that is, $(PT)^{-1}HPT = H$ one finds that this requirement is equivalent as stating that the complex potential function has a real part that is even while the complex part is odd under $\mathbf{r}\rightarrow - \mathbf{r}$. These authors studied the eigenvalues of the PT-symmetric Hamiltonian $H = p^2 - (ix)^\varepsilon$ as a function of the real parameter $\varepsilon$ and found a symmetry breaking point, depending on this parameter. More specifically, when $\varepsilon \geq 2$ the eigenvalues are real, positive and discrete. On the other hand, in the so-called broken region, i.e., $\varepsilon < 2$, the eigenvalues switch from real to complex values, and now appear in complex pairs of the form $a\pm ib$. Thus, even though the Hamiltonian is PT-symmetric for every $\varepsilon$, below the symmetry breaking point, or the exceptional point, the spectrum of the PT-symmetric Hamiltonian becomes complex \cite{Heiss_2004}. More specifically, the complex conjugate eigenvalue pairs represent non-equilibrium systems with exponential amplification and dissipation, therefore they represent open systems \cite{Benderbook,rotter2009}. However, below the PT phase transition, the system behaves like a closed system due to the all-real energy spectrum and by a new definition of the inner product in the Hilbert space \cite{Bender2002}. It was later shown in a series of papers published by Mostafazadeh that PT-symmetric Hamiltonians actually belong to a subclass of a more general set of operators named pseudo-Hermitian Hamiltonians \cite{Mostafazadeh20021,Mostafazadeh20022,Mostafazadeh20023}. 

As waves always behave in a similar way, the notion of exceptional points and PT symmetry has revealed itself to be quite useful in various areas of physics, such as electromagnetic invisibility cloaks \cite{zhu2013}, acoustic scattering \cite{zhu2014} and metasurfaces \cite{sounas2015}, to cite a few. Owing to the analogy between the time-dependent Schr\"odinger equation and the scalar paraxial wave equation, derived from Maxwell's relations, an earlier experimental verification of a PT-symmetric physical model was performed within the context of optics, by using waves with a high degree of coherence, i.e., laser beams, interacting with coupled waveguides \cite{Ruter2010,Christodoulides2007,guo2009}. The potential energy $V(x)$ of a one-dimensional quantum system is interpreted as the complex refractive index $n(x)$ whose imaginary part represents regions in space of gain and loss, distributed in a balanced way. Many unexpected optical effects were unraveled by the incidence of monochromatic radiation into PT-symmetric media as for example in ultra-sensitive sensors \cite{Ren2017}. The huge impact that PT-symmetric ideas have printed upon laser science and technology may be evaluated by the number of emerging areas that there has been since. One remarkable example is the idea of a laser-absorber device \cite{longhi2010, chong2011}. A recent comprehensible review on the applicability of PT optics and non-Hermitian photonics is given in \cite{Longhi2018}. Nowadays researchers have also realized that the application of the ideas of PT-symmetry and exceptional points to optical systems has become an important area of research in integrated photonics \cite{ozdemir2019}. 

However, there is one particular important aspect of the optical fields that remained untouched by the new concepts of PT-symmetric quantum mechanics. The fluctuations of the incident radiation field and their influence on the spectral density of the field, on scattering, have been completely ignored in all studies, at the best of our knowledge. It is well known that complete spatial or temporal coherence of radiation does not exist and thus, a theory that includes the random nature of the incident field, should be of some importance to any reader that deals with the wave nature of radiation \cite{wolfbook}. In this way, the scattering of partially coherent radiation by a deterministic periodic medium has been addressed in \cite{Dusek1995,Wolf2013} within the formalism of classical coherence theory. Therefore, considering the huge success of PT-symmetry in practical applications in optics, and furthermore, that field fluctuations are ubiquitous in real photonic systems, in this work we describe a first attempt to elucidate some of the intriguing aspects involving the scattering of a partially coherent radiation field, by a PT-symmetric material with localized regions of gain and loss. Earlier treatments of scattering of coherent radiation by PT-symmetric structures are available \cite{Hurwitz2016,miri2016}. To this end, in section II we describe the model of a PT-structure, based on a pair of particles, representing localized gain and loss materials. We then apply classical coherence theory, based on the Born approximation, to obtain the spectral density changes of a partially coherent field propagating in the far zone, due to the scattering by the two point scatterers. Section III is devoted to the detailed analysis of the spectral density obtained in Section II, providing more clarity into the origins of the spectral changes. Finally, in Section IV, we conclude and discuss further developments.

\section{Physical model}

Let us begin by considering the formalism of classical coherence theory and suppose that all statistical quantities are essentially stationary. We wish to determine the far-field spectral density $S^{(\infty)}(\mathbf{r},\omega)$, at position $\mathbf{r}$ and angular frequency $\omega$, of a partially coherent radiation field that has been scattered by a localized non-Hermitian PT-symmetric structure.

\begin{figure}[hbt]
	\centering
	\includegraphics[width=0.7\linewidth]{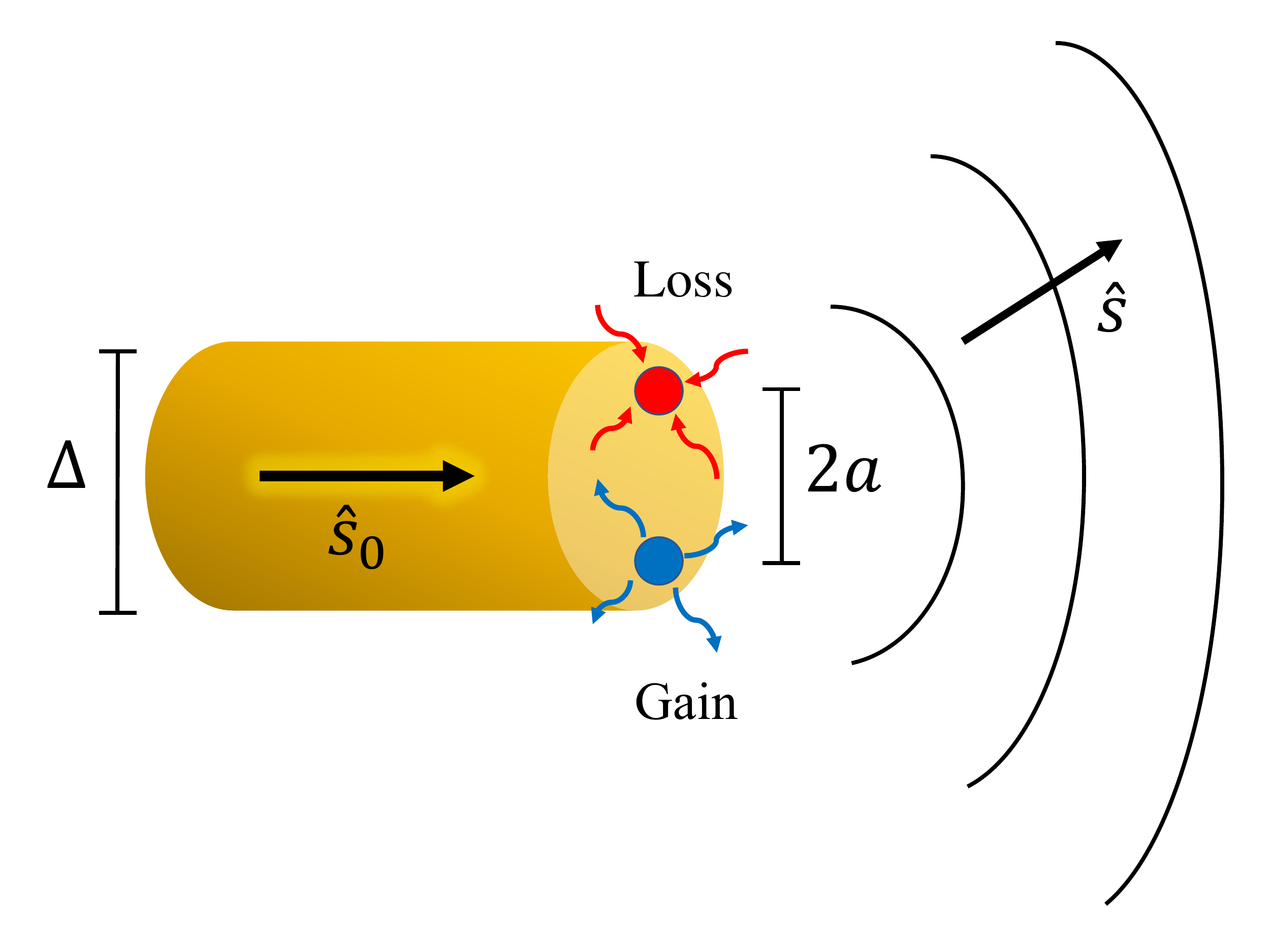}
	\caption{(Color online) The model consists of two point scatterers separated by distance $2a$, one having loss $+i\gamma$ (in red) at position $x = a$ and the other having gain $-i\gamma$ (in blue) at position $x = -a$. They are situated in the optical path of a partially coherent beam with coherence length $\Delta$. The incident direction of the beam is represented by the unit vector $\hat{s}_0$ and the scattered far-field direction is represented by the unit radial vector $\hat{s}$.}
	\label{fig1}
\end{figure}

Let us then define the cross-spectral power density of the incident wavefield
$\{ u^{(i)}(\mathbf{r},\omega) \}$,
by $W^{(i)}(\mathbf{r}_1,\mathbf{r}_2,\omega) = \langle 
u^{(i)}(\mathbf{r}_1,\omega)^{*}u^{(i)}(\mathbf{r}_2,\omega) \rangle$. Here,  
the brackets denote the average over a statistical ensemble of monochromatic 
realizations of the incident field. Analogously, the cross-spectral density of the scattered wavefield $u^{(s)}(\mathbf{r},\omega)$ is defined as $W^{(s)}(\mathbf{r}_1,\mathbf{r}_2,\omega) = \langle u^{(s)}(\mathbf{r}_1,\omega)^{*}u^{(s)}(\mathbf{r}_2,\omega) \rangle$ and the two are related by the integral \cite{wolfbook}
\begin{multline}\label{wswi}
	W^{(s)}(\mathbf{r}_1,\mathbf{r}_2,\omega) = \int_V \int_V W^{(i)}(\mathbf{r}_1',\mathbf{r}_2',\omega)F^* (\mathbf{r}_1',\omega) 
	F(\mathbf{r}_2',
	\omega) \\
	\times G^*(|\mathbf{r}_1-\mathbf{r}_1'|,\omega) 
	G(|\mathbf{r}_2-\mathbf{r}_2'|,
	\omega)d^3 r_1' d^3 r_2',
\end{multline}
where $F(\mathbf{r}, \omega)$ is the potential function proportional 
to the refractive index of the material and $G$ represents the Green's function of the system. We will consider a localized scatterer, as depicted in Figure \ref{fig1}, described by the function
\begin{equation}\label{Fr}
    F(\mathbf{r},\omega) = \delta(y)\delta(z)[ (\sigma +i\gamma)\delta(x-a) +(\sigma -i\gamma)\delta(x+a)],
\end{equation}
with $\sigma$ and $\gamma$ being real positive parameters for each individual scatterer, which can be $\omega$-dependent. Since $\gamma$ is associated with the imaginary part of the potential, it describes the non-Hermitian properties of the material. Thus, the scatterer located at $x = a$ has intrisinc loss ($+i\gamma$) while the scatterer located at $x = -a$ has intrinsic gain $-i\gamma$ (see Figure \ref{fig1}). One can easily 
verify that the above material function satisfies the PT-symmetry condition: 
\begin{equation}
(PT)^{-1}F(\mathbf{r},\omega)PT = F(-\mathbf{r},\omega)^* = F(\mathbf{r},\omega).
\end{equation}
The outgoing free-space Green's function is given by
\begin{equation*}
	G(|\mathbf{r}-\mathbf{r}'|,\omega) = \frac{e^{ik|\mathbf{r}-\mathbf{r}'|}}{|\mathbf{r}-\mathbf{r}'|}
\end{equation*}
where $k=\frac{\omega}{c}$ with $c$ being the speed of light in vacuum. In deriving Equation \eqref{wswi}, the first-order Born approximation was used, so that the scattered spectral density now reads \cite{wolfbook} 
\begin{equation}
S^{(s)}(\mathbf{r},\omega) = W^{(s)}(\mathbf{r},\mathbf{r},\omega).
\end{equation}
Furthermore, as we are interested in the fields in the far-zone region where $k|\mathbf{r}| \rightarrow \infty$, the Green's function becomes 
$G \sim \frac{e^{ikr}}{r}e^{-ik\hat{s}\cdot\mathbf{r}'}$, where $\hat{s} = \frac{\mathbf{r}}{|\mathbf{r}|}$ and $r = |\mathbf{r}|$. Assuming these approximations, the far-field spectral density $S^{(\infty)}(\mathbf{r},\omega)$ is finally obtained 
\begin{multline}\label{Sinf}
	S^{(\infty)}(\hat{s},\omega) = \frac{S^{(i)}(\omega)}{r^2}\int_V\int_V \mu^{(i)}(\mathbf{r}_1,\mathbf{r}_2,\omega)	F^* (\mathbf{r}_1,\omega) \\
	\times F(\mathbf{r}_2,\omega) 
	e^{-ik\hat{s}\cdot(\mathbf{r}_2-\mathbf{r}_1) } d^3 r_1 d^3 r_2,
\end{multline}
where we have made the assumption that the spectral density of the incident wavefield is independent of position and it is represented by $S^{(i)}(\omega) = W^{(i)}(\mathbf{r},\mathbf{r},\omega)$. The function $\mu^{(i)}(\mathbf{r}_1,\mathbf{r}_2,\omega)$ is the normalized cross-spectral density, also known as the spectral degree of coherence. Since our objective is to construct a simple physical model suitable to describe the essential characteristics of the scattering process, we must choose the spectral degree of coherence and the scattering potential in such a way that the integral in \eqref{Sinf} may be analytically solved. With the choice of $F(\mathbf{r},\omega)$ given by Equation \eqref{Fr}, we now assume a partially coherent incident homogeneous 
wavefield of infinite width, whose spectral degree of coherence is given by,   
\begin{equation}\label{mui}
\mu^{(i)}(\mathbf{r}_1,\mathbf{r}_2,\omega) = e^{-\frac{(\boldsymbol{\rho}_1 - \boldsymbol{\rho}_2)^2}{2\Delta^2}}e^{ik\hat{s}_0\cdot (\mathbf{r}_2 - \mathbf{r}_1)},
\end{equation}
where $\hat{s}_0$ is the incident field direction, $\hat{s}_0 \cdot \boldsymbol{\rho}_{1,2} = 0$ with $\boldsymbol{\rho}$ denoting a transverse position vector and $\Delta$ is the coherence length of the beam which characterizes the distance between a pair of points in the wavefield such that correlation exists. Equation 
\eqref{mui} defines the coherence properties of the incident field,
whose transverse points are correlated in a Gaussian fashion: when 
$\Delta \rightarrow \infty$, the wavefield is fully spatially 
coherent as $|\mu^{(i)}(\mathbf{r}_1,\mathbf{r}_2,\omega)| = 1$, meaning that any point in the beam is correlated with every other. After substituting Equations \eqref{Fr} and \eqref{mui} into Equation \eqref{Sinf} we obtain the far zone 
spectral density
\begin{multline}\label{Smain}
	S^{(\infty)}(\hat{s},\omega) = S^{(i)}(\omega)\frac{2(\sigma^2 + 
	\gamma^2)}{r^2}
	\left\{ 1 + e^{ -\frac{2a^2}{\Delta^2} } 
	\right. \\ 
	\left.  \times \left[ \left( \frac{\sigma^2-\gamma^2}
	{\sigma^2+\gamma^2} 
	\right)
	\cos(2kas_x) + \left(  \frac{2\sigma\gamma}{\sigma^2 + 
	\gamma^2} 
	\right)\sin(2kas_x) \right] \right\},
\end{multline}
where $s_x = \sin\theta\cos\phi$ is the $x$ component of the radial unit vector $\hat{s}$ and we have assumed that $\hat{s}_0 = \hat{z}$, 
i.e., the incident wavefield is propagating in the positive $z$ direction. 
Equation \eqref{Smain} is the central equation of this work. It provides 
the spectral density of a partially coherent wavefield that has 
been scattered by a PT-symmetric material described by the material 
function of Equation \eqref{Fr}.

\section{Far-zone non-Hermitian scattered Spectral Density}

To grasp the physics behind Equation \eqref{Smain}, let us consider some special cases. First, when the structure is Hermitian, $\gamma = 0$ so that
\begin{equation}\label{SH}
S^{(\infty)}_{H}(\hat{s},\omega) = S^{(i)}(\omega)\frac{2\sigma^2}{r^2}\left[ 1+e^{-\frac{2a^2}{\Delta^2}}\cos(2kas_x) \right].
\end{equation}
Equation \eqref{SH} describes the spectral density in a passive system with a real refractive index. It is seen from this relation that the scattered spectrum may differ in general from the incident one depending on its particular state of coherence. When $\Delta \gg 2a$, that is, when the two scatterers are localized in a region within the coherence length of the incident field, an interference pattern is obtained. Otherwise, i.e, when $\Delta \le 2a$, that is, when the scatteres are located in a region approximately equal or larger than the coherence length of the incident wavefield, the interference pattern loses its visibility and eventually vanishes, a very well-known result \cite{wolfbook}. 

\begin{figure}[hbt]
	\centering
	\includegraphics[width=1\linewidth]{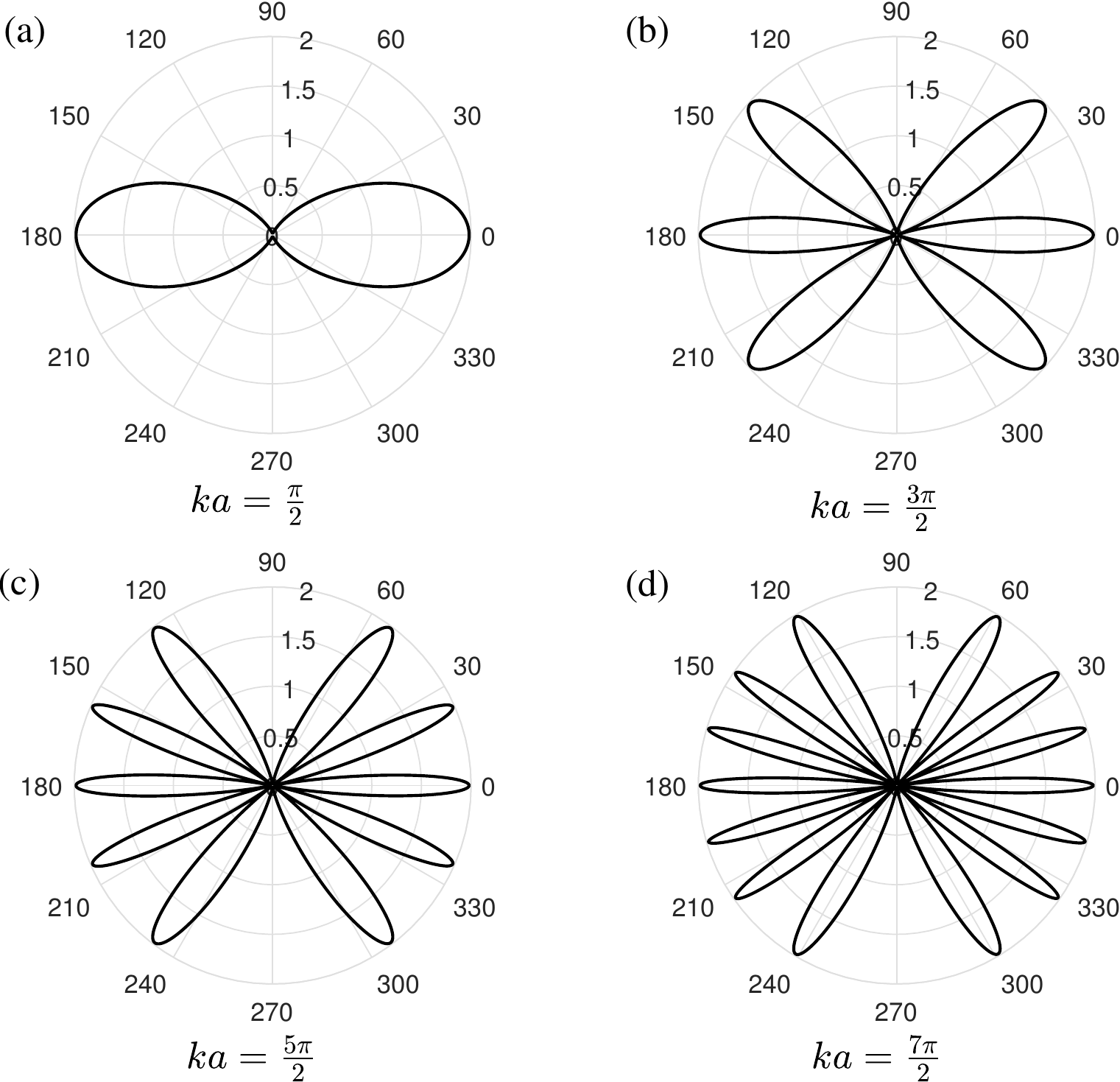}
	\caption{Hermitian spectral density $r^2 S_{H}^{(\infty)}(\hat{s},\omega)/2\sigma^2 S^{(i)}(\omega)$ generated by a Hermitian two-point scatterers with separation distance $a_n = \frac{n\pi}{2k}$ with (a) $n = 1$, (b) $n = 3$, (c) $n = 5$ and (d) $n = 7$. When the distance between the sources are given by $a_n$, the system emits no radiation at $\theta = \frac{\pi}{2}$ and $\theta = \frac{3\pi }{2}$ as can be deduced from the general expression \eqref{Smain} in the high coherence limit $\Delta \rightarrow \infty$. }
	\label{fig2}
\end{figure}

\begin{figure}[hbt]
	\centering
	\includegraphics[width=0.7\linewidth]{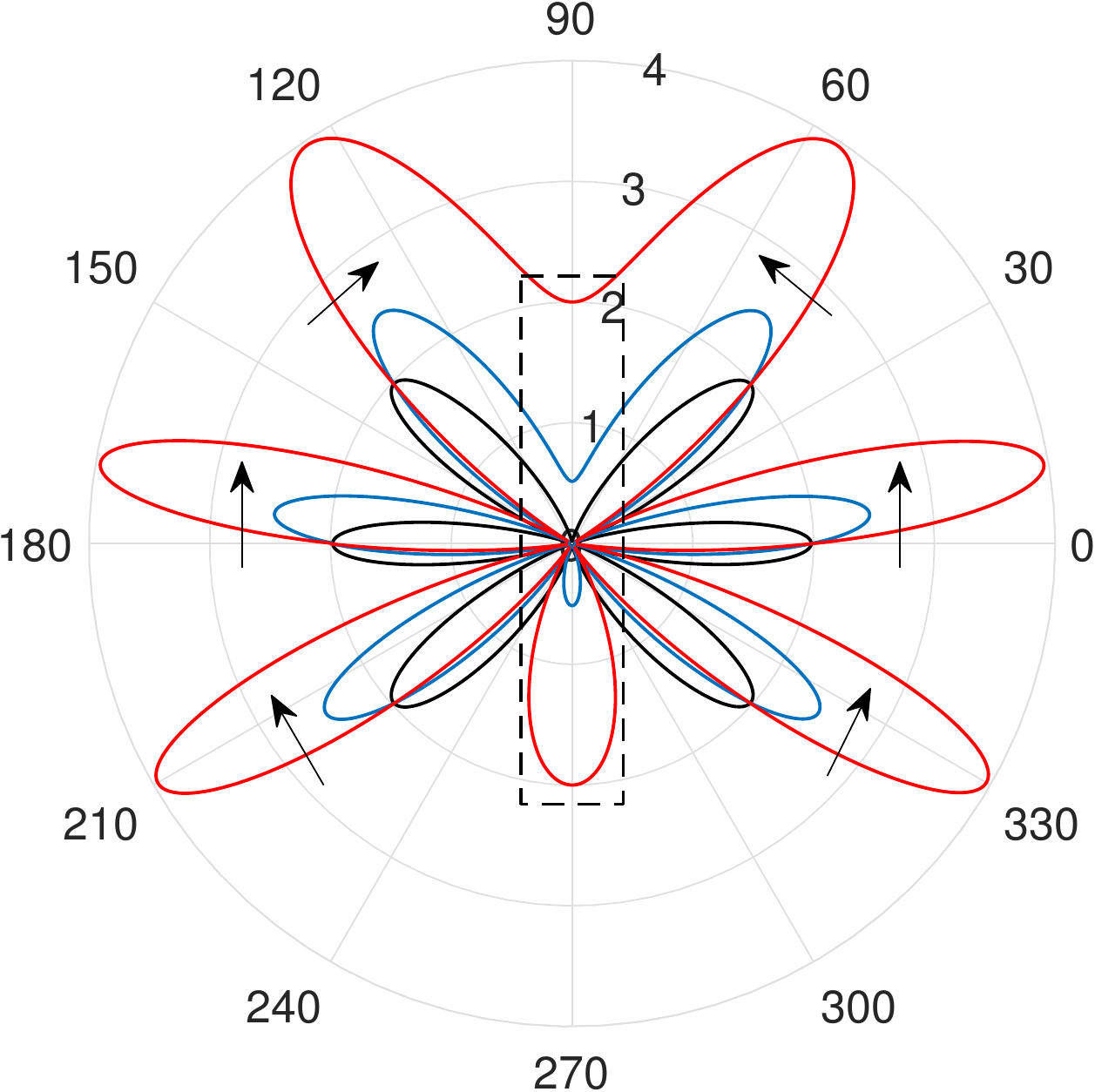}
	\caption{(Color online) Normalized spectral density $r^2 S^{(\infty)}(\hat{s},\omega)/2S^{(i)}(\omega)$ as a function of $\theta$. The range $\theta > \pi $ simply means that $\phi = \pi$. Black line: Hermitian scattering described by Equation \eqref{SH}. Blue line: Non-Hermitian, PT-symmetric, spectral density calculated using Equation \eqref{Smain} with 
	$\gamma = 0.5$. Red line: Non-Hermitian spectral density with $\gamma = 1$, obtained from \eqref{Smain}. In this plot, $ka = \frac{3\pi}{2}$	and $k\Delta = 10ka$, meaning that the scatterer is 
	situated in a region where the wavefield vibrations are highly correlated. The black arrows indicate the sense of rotation of the interference fringes as the non-Hermitian parameter $\gamma$ increases. Inside the dotted box, non-negligible radiation appears which are not present in its Hermitian counterpart. The value $\sigma = 1$ was assumed in all these plots.}
	\label{fig3}
\end{figure}

\begin{figure}[hbt]
	\centering
	\includegraphics[width=0.7\linewidth]{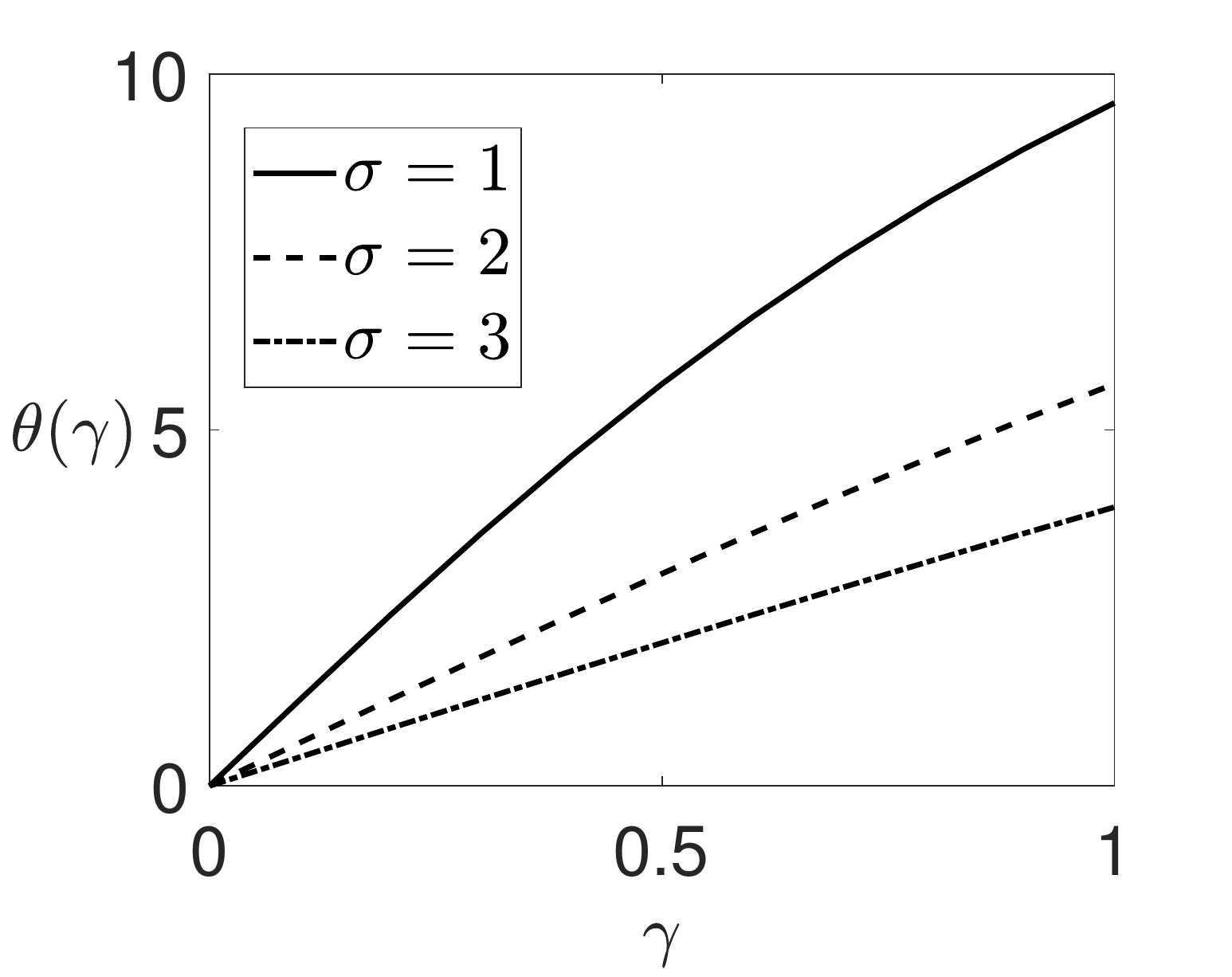}
	\caption{ Angle (in degrees) of the maximum emitted radiation in the direction $\theta = 0$ (when $\gamma = 0$ initially) as a function of the non-Hermitian parameter $\gamma$. The variation of the angular direction is given by Equation \eqref{dev}. In this figure we have used $ka = \frac{3\pi}{2}$, as in Figure \ref{fig3}. The deflection is independent of the coherence length $\Delta$.}
	\label{fig4}
\end{figure}

We see from Equation \eqref{SH} that if the distance between the scatterers satisfy $2ka_n = n\pi$ with $n$ odd, there is no radiation scattered into the $\theta = \frac{\pi}{2}$ and $\theta = \frac{3\pi}{2}$ directions if $\Delta \gg 2a$ (high coherence limit). Figure \ref{fig2} illustrates the polar plot for the Hermitian spectral density with $n =$ 1, 3, 5 and 7 in the incident plane defined by $\phi = 0$ ($s_x = \sin\theta$) and $\phi = \pi$ ($s_x = -\sin\theta$). It is clearly seen in these pictures the inhibition of radiation for the particular values of $\theta = \frac{\pi}{2},\frac{3\pi}{2}$. We will be mainly interested in these two directions because it is around these angles that the spectral density changes more drastically. Note that these directions are in the line of the two scatterers, i.e., they are perpendicular to the direction of the incident radiation. So, we conclude that a passive, Hermitian, configuration does not radiate perpendicularly to the incident direction for these particular values of $a_n$. Note also that the number of intensity maxima in the interference pattern is twice the discrete index $n$ and that the spectral density is symmetric under $\theta \rightarrow \pi-\theta$ ($0\leq\theta \leq \pi$).
Let us consider $2ka = 3\pi$ fixed from now on and investigate the complex extension of the spectral density for $\gamma > 0$. Figure \ref{fig3} shows the non-Hermitian spectral density given by \eqref{Smain} for three values of $\gamma$. The black line in Figure \ref{fig3} is the Hermitian configuration with $\gamma = 0$ which is also shown in part ($b$) of Figure \ref{fig2}. The blue line indicates the non-Hermitian, PT-symmetric, spectral density with $\gamma = 0.5$ and the red line was obtained with $\gamma = 1.0$. All three plots were generated by fixing $\sigma = 1$ and $k\Delta = 10ka$. This implies that the coherence length $L_c = \Delta$ and the distance between scatterers $L_s = 2a$ are in ratio $\frac{L_c}{L_s} = 5$ so that the scatterers are inside the coherence length of the wavefield and therefore the wavefield vibrations in the region around the scatterers are highly correlated. There are two main effects discernible in this non-Hermitian situation. First, as Figure \ref{fig3} clearly indicates, the intensity maxima rotate in such a way that they tend towards the angle $\theta = \frac{\pi}{2}$ as $\gamma$ increases. This is indicated in the figure by black arrows pointing to the direction of rotation. The second effect is the asymmetrical appearance of scattered radiation at angles $\theta = \frac{\pi}{2}$ and $\theta = \frac{3\pi}{2}$ that was inhibited in the Hermitian configuration. We highlight this new radiation direction by a dotted box in Figure \ref{fig3}. The exact angular rotation is easily calculated from the general relation \eqref{Smain}. The spectral density maximum at $\theta = 0$ (when $\gamma = 0$), for example, deviates from $\theta = 0$ according to the relation
\begin{equation}\label{dev}
    \theta(\gamma) = \sin^{-1}\left[ \frac{1}{2ka}\tan^{-1}\left( \frac{2\sigma\gamma}{\sigma^2-\gamma^2} \right) \right]
\end{equation}
which is plotted in Figure \ref{fig4} for three values of $\sigma$. Equation \eqref{dev} is valid for $0\leq \gamma \leq 1$ and one can see from Figure \ref{fig4} that the deviation is almost linear in this interval. Also, larger values of $\sigma$ lead to smaller deviation angles from zero for a fixed $\gamma$. When $\sigma = \gamma$ we have $\theta(\gamma) = \sin^{-1}\left(\frac{\pi}{4ka}\right)$ which depends only on the separation distance between the scatterers.

\begin{figure}[hbt]
	\centering
	\includegraphics[width=0.7\linewidth]{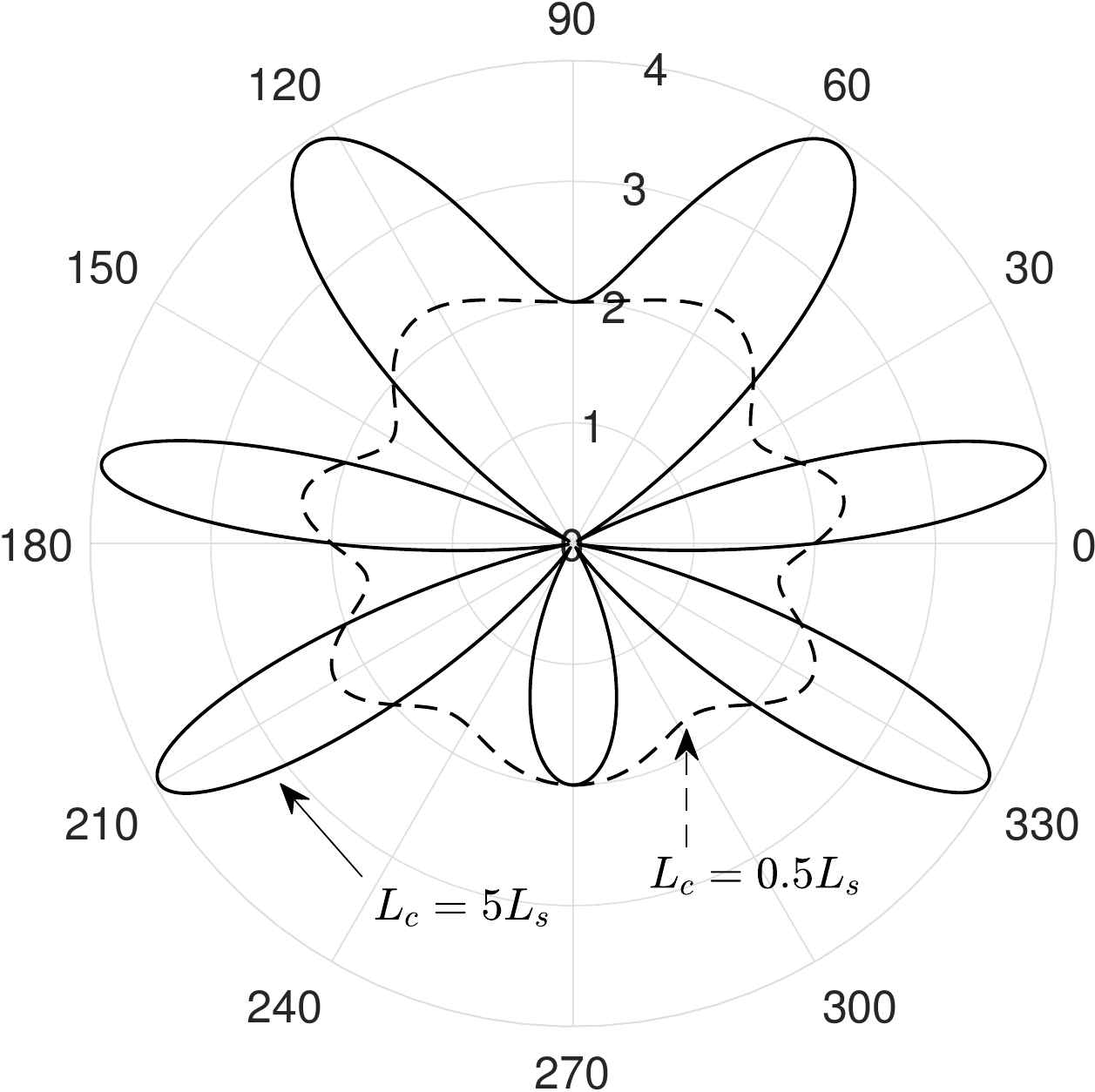}
	\caption{Non-Hermitian spectral density of partially coherent diffracted wavefields. The continuous line represents the configuration where the two scatterers are situated in a region of high coherence, $L_c = 5L_s$. The dotted line is plotted with $L_c = \frac{L_s}{2}$ which indicates that the wavefield around each scatterer is poorly coherent. The coherence properties of the wavefield thus inhibits the asymmetrical directional character typical of non-Hermitian systems.}
	\label{fig5}
\end{figure}

This asymmetric directional character of the scattered radiation present in non-Hermitian optical configurations has been the subject of intense research since the first papers dealing with beam propagation in PT-symmetric lattices or coupled waveguides \cite{Makris2008,Gao2018,Ruter2010,Zhu2016,shui2018}. Here we see that if the classical coherence of a wavefield is taken into account, one can destroy these important and unusual non-Hermitian properties. Figure \ref{fig5} illustrates this aspect by showing the spectral density in two different configurations. In the first case, the two scatterers are situated in a region where the optical wavefield is highly coherent, $L_c = 5 L_s$, meaning that the field fluctuations around the two scatterers are correlated according to Equation \eqref{mui}. This situation is represented by the continuous line in Figure \ref{fig5}. The other situation assumes that the wavefield fluctuations are not correlated in the region around the scatterers, $L_c = \frac{L_s}{2}$. In this more realistic scenario, the asymmetric character of the emitted radiation is suppressed by the coherence properties of the wavefield and, as a consequence, the material begins to radiate into previously forbidden regions. Therefore, one finds more possibilities to manipulate the character of the emitted radiation by tuning the coherence properties of the incident wavefield, without changing the geometry of the system.

\begin{figure}[hbt]
	\centering
	\includegraphics[width=0.8\linewidth]{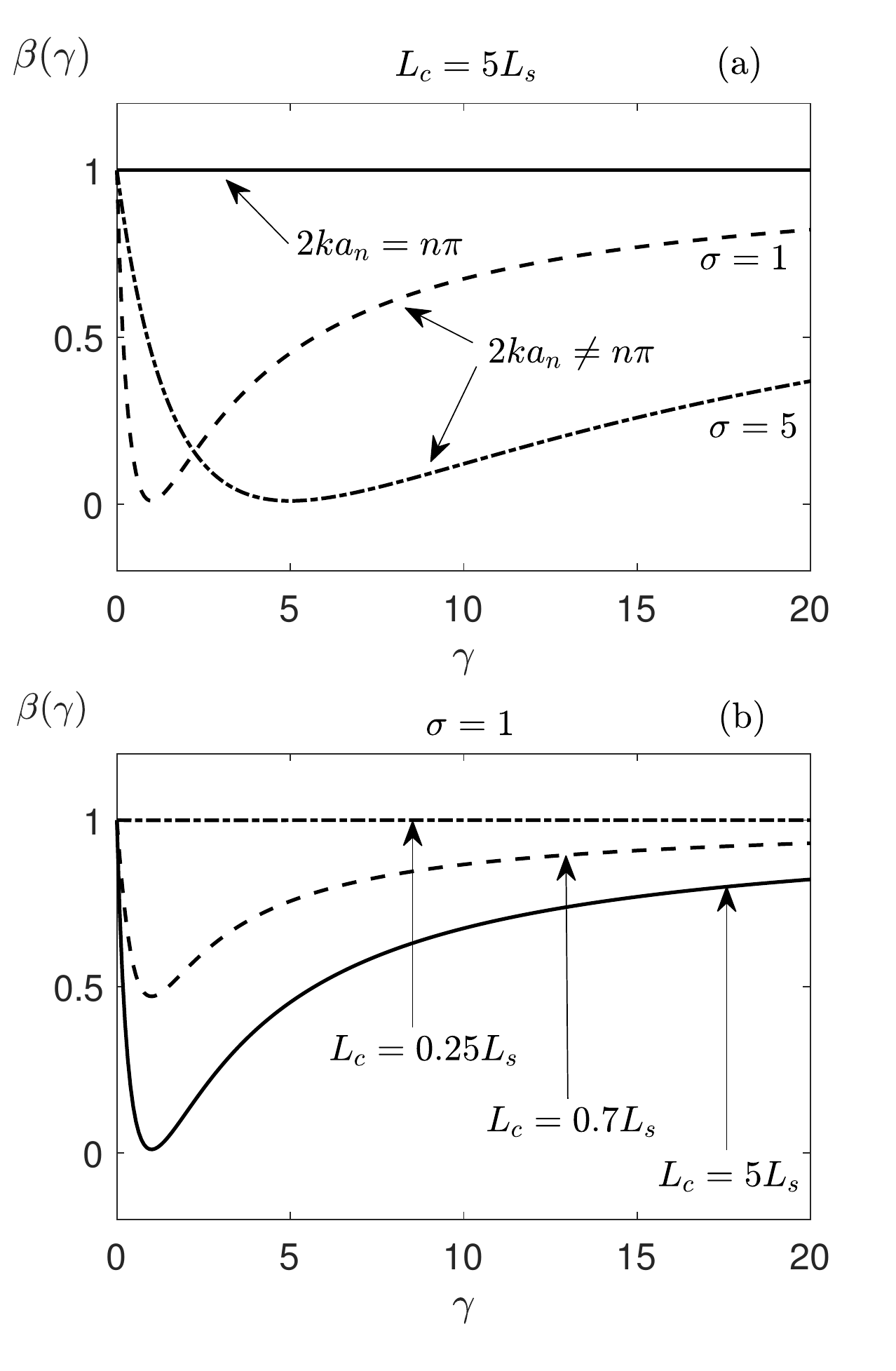}
	\caption{(a) Plot of $\beta(\gamma)$, defined by Equation \eqref{beta}, as a function of $\gamma$ for fixed $L_c = 5L_s$ (high coherence situation). The curves show the situation where $2ka_n = n\pi$ (continuous line), $\sigma = 1$ (dotted line) and $\sigma = 5$ (dashed-dotted line). In the last two cases, the minimum value of $\beta$ is achieved when $\gamma = \sigma$. (b) Plot of $\beta(\gamma)$ for fixed $\sigma = 1$ and $L_c = 5L_s$ (continuous line), $L_c = 0.25L_s$ (dotted line) and $L_c = 0.25L_s$ (dashed-dotted line).}
	\label{fig6}
\end{figure}

A nontrivial relation becomes evident by comparing the value of the spectral density at two different directions perpendicular to the incident direction. We define the quantity 
\begin{equation}\label{beta}
    \beta(\gamma) = \frac{S^{(\infty)}(\theta = \frac{\pi}{2},\phi = 0)}{S^{(\infty)}(\theta = \frac{\pi}{2},\phi = \pi)}
\end{equation}
as a measure of the ratio between the value of the spectral density at $(\theta = \frac{\pi}{2},\phi = 0)$ and $(\theta = \frac{\pi}{2},\phi = \pi)$ as we change the non-Hermitian parameter $\gamma$. The interesting thing is that we obtain two completely different behaviors depending on the distance between the scatterers. If this distance satisfies the condition assumed in the previous analysis, i.e., $2a_n = \frac{n\pi}{k}$, then $\beta(\gamma) = 1$ \textit{independent of} $\gamma$, $\Delta$ and $\sigma$, as can be easily calculated by using the relations \eqref{beta} and \eqref{Smain}. This effect can be seen in Figures \ref{fig3} and \ref{fig5}. A plot of $\beta(\gamma)$ in this situation would give us a horizontal line starting at $\beta = 1$ and this implies that the magnitudes of the emitted spectral intensity perpendicular to the incident direction are equal. However, if $2a_n \neq \frac{n\pi}{k}$ the function $\beta(\gamma)$ exhibits a minimum value at $\gamma = \sigma$. Part (a) of Figure \ref{fig6} shows the plot of $\beta(\gamma)$ for fixed $L_c = 5L_s$ (high coherence situation). If 
$2ka_n = n\pi$ the plot is the line $\beta = 1$ and if $2ka_n \neq n\pi$, $\beta$ shows a \textit{minimum} at $\gamma = 1$ (dashed line) and $\gamma = 5$ (dashed-dotted line). The value of $\beta$ at both minima in part (a) are equal: $\beta(\gamma = \sigma = 1) = \beta(\gamma=\sigma=5)$. This 
is because the value of $\beta(\gamma)$ at $\gamma = \sigma$ is independent 
of $\sigma$:
\begin{equation}\label{betasigmagama}
    \beta(\gamma = \sigma) = \frac{1 + e^{-2\frac{a^2}{\Delta^2}}\sin(2ka)}{1 - e^{-2\frac{a^2}{\Delta^2}}\sin(2ka)}.
\end{equation}
In the limit $\gamma\rightarrow +\infty$, 
\begin{equation}
\beta(\gamma) \sim \frac{1+e^{-2\frac{a^2}{\Delta^2}}\left[\cos(2ka)+\frac{2\sigma}{\gamma}\sin(2ka)\right]}{1+e^{-2\frac{a^2}{\Delta^2}}\left[\cos(2ka)-\frac{2\sigma}{\gamma}\sin(2ka)\right]},
\end{equation}
and so it approaches $1$ for fixed $\sigma$ and $a$. Part (b) of Figure \ref{fig6} is considered at $\sigma = 1$ fixed. The objective of this plot is to show the effect of coherence on the function $\beta$. When the coherence length $\Delta$ of the wavefield is larger than the distance between the scatterers, the function $\beta$ has a minimum at $\beta = \sigma$ with the value of $\beta$ given by Equation \eqref{betasigmagama} and so the effect of increasing $L_c$ compared to $L_s$ is to increase the minimum value of $\beta$ until it gets equal to unity at the low coherence regime $L_c \ll L$. A zero value of $\beta(\gamma)$ corresponds to radiation being emitted mainly around $\theta = \frac{3\pi}{2}$, the direction more closely related to the scatterer with gain (located at $x = -a$). 

To gain a deeper understanding on the interplay between PT-symmetry
and classical correlations during scattering processes, it is useful to 
calculate the cross-spectral density function $W^{(s)}(\mathbf{r}_1,\mathbf{r}_2,\omega)$ of the scattered radiation field directly from Equation \eqref{wswi} with the potential function \eqref{Fr} to obtain
\begin{align}\label{wsr1r2}
 W^{(s)}(\mathbf{r}_1,\mathbf{r}_2,\omega) &= \frac{2(\sigma^2+\gamma^2)e^{ik(r_2-r_1)}}{r_1 r_2} \nonumber\\
 &\times \left\{ W_{S} \cos[ka(s_{1x}-s_{2x})] \right.\nonumber\\
 &+ \left. {} \left(\frac{\sigma^2-\gamma^2}{\sigma^2+\gamma^2}\right)W_{C}\cos[ka(s_{1x}+s_{2x})] \right. \nonumber \\
 &+ \left. {} \left(\frac{2\sigma\gamma}{\sigma^2+\gamma^2}\right)W_{C}\sin[ka(s_{1x}+s_{2x})] \right\}
\end{align}
where $s_{jx}$ is the $x$-component of the unit vector $\hat{s}_j = \mathbf{r}_j/|\mathbf{r}_j|$ for $i \in \{ 1,2 \}$ and
\begin{equation}
\begin{aligned}
  W_S &= W^{(i)}(x_1 = \pm a,x_2 = \pm a,\omega), \\
  W_C &= W^{(i)}(x_1 = \pm a,x_2 = \mp a,\omega), 
\end{aligned}
\end{equation}
are the cross-spectral densities of the \textit{incident} field 
at the positions of the two scatterers with $W_S$ representing self-correlation and $W_C$, cross-correlation. The presence of $\gamma$ 
in Equation \eqref{wsr1r2} suggests that the two-point correlation
function depend crucially on the gain/loss properties of the material, changing its amplitude and/or phase according to the specific values of $\sigma$, $\gamma$. Thus, one might conclude that a PT-symmetric medium
may be used to manipulate the correlations properties of the 
scattered optical field. 

Generally speaking, when one is dealing with the Schr\"{o}dinger equation with a PT-symmetric potential satisfying $V(x) = V(-x)^{*}$, the time evolution of a state is crucially dependent on the spectral properties of the Hamiltonian. It is expected that if the Hamiltonian is dependent upon a parameter, $H(q)$, with $q$ a real number, then there exists a critical value $q_c$ such that the eigenvalues of $H$ are real for $q < q_c$ and form complex conjugated pairs for $q > q_c$, as discussed in the introduction section. The time evolution of a physical state $\psi$ is unitary under a $CPT$ inner product and this is only possible in the regime where all eigenvalues are real \cite{Bender2002}. However, in the present scattering model, we have found no indication of any divergences for any value of $\gamma$ and $\sigma$. One might conjecture that these non-unitary effects are not apparent considering 
that our results are obtained from a scattering approach. The usual symmetry breaking phenomenon does become mathematically noticeable when the results are obtained within a scattering matrix approach \cite{alu}. Here, we find no abrupt change 
in the spectral functions at all when 
crossing the symmetry breaking point $\sigma = \gamma$ and we attribute it to the fact that PT-symmetric systems exhibit differences from scattering problems as they are subjected to different constraints. The question on the 
existence and behavior of exceptional points in classical optical 
coherence theory is still an open question worth
to be pursued. However, we do not attempt to give a satisfactory answer in the present paper. We hope however that this manuscript may provide a first motivation for the initial studies on the interplay of 
PT symmetry and classical coherence effects, which are still lacking in the current literature.

\section{Conclusions}

In conclusion, we have found dramatic non-Hermitian changes occurring in the spectral density and in the cross spectral function of a 
scattered wavefield by a PT-symmetric localized material, considering the coherence properties of the source. We have obtained explicit analytic expressions for these  spectral properties under the Born approximation and analyzed many limiting cases to grasp the physics of the interplay between PT-symmetry and partial coherence. In summary, the coherence length of the partially coherent incident beam may alter the asymmetric directional character of the emitted radiation, an effect that has not been previously reported. Furthermore, the cross spectral 
correlation exhibits a crucial dependence on the gain/loss spatial
distribution. This feature might be very important for application 
purposes due to the possibilities it provides to manipulate the 
nuisances of fluctuations by non-Hermitian structures.  As in real life, optical fields are produced by partially coherent sources, and considering the ability of non-Hermitian structures to control and manipulate light transport, we believe that the study on the interaction of PT symmetric structures with partially coherent light should reveal itself quite fruitful with many effects waiting to be explored. 

\begin{acknowledgments}
We wish to acknowledge the partial financial support of the Brazilian agencies, CNPq and FAPEAL
\end{acknowledgments}
%

%

%

%




\end{document}